\documentclass[prb,nofootinbib,twocolumn,superscriptaddress]{revtex4} 

\usepackage{graphicx}
\usepackage{dcolumn}
\usepackage{bm}
\usepackage{threeparttable}
\usepackage{times}
\usepackage{mathptmx}
\usepackage{lscape}
\usepackage{natbib}
\usepackage{amsmath}
\usepackage{amssymb}
\usepackage{braket}
\usepackage{comment}
\usepackage{color}

\def\degree{\kern-.2em\r{}\kern-.3em}

\begin{document}

\title{ Trends in Gibbs States for Thermodynamics of Canonical Nonlinearity  }

\author{Ryu Tomitaka}
\affiliation{
Department of Materials Science and Engineering,  Kyoto University, Sakyo, Kyoto 606-8501, Japan\\
}%

\author{Koretaka Yuge}
\affiliation{
Department of Materials Science and Engineering,  Kyoto University, Sakyo, Kyoto 606-8501, Japan\\
}%

\begin{abstract}
{ When we consider canonical average for classical discrete systems under constant composition (specifically, substitutional alloys) as a map $\phi$ from a set of many-body interatomic interactions to that of microscopic configuration in thermodynamic equilibrium, $\phi$ generally exhibits complicated nonlinearity. The nonlinearity has recently been amply studied in terms of configurational geometry, measured by vector field and Kullback-Leibler divergence, whose individual concepts are further unified through stochastic thermodynamic transformation: We call this procedure as ``Themodynamics of canonical nonlinearity (TCN)''.  Although TCN can reveal the nonlinear character across multiple configurations through thermodynamic functions, the essential role for Gibbs states (GBS) in terms of the nonlinearity is still totally unclear. We here tackle this problem, and reveal the characteristic roles of the GBS: (i) Concrete expression of the GBS is derived, (ii) strong correlation between GBSs and averaged nonlinearity over configurations for moled systems is found, and (iii) GBS-based bounds for averaged nonlinearity at specific conditions is derived.

}
\end{abstract}


\maketitle

\section{Introduction}
When we consider classical discrete systems under constant composition on given lattice, canonical average 
$\Braket{\quad}_{Z}$ provides expectation for configuration in thermodynamic equilibrium under prepared coordination of $q_{1},\cdots, q_{f}$ , namely, 
\begin{eqnarray}
\label{eq:can}
\Braket{ q_{p}}_{Z} = Z^{-1} \sum_{i} q_{p}^{\left( i \right)} \exp \left( -\beta U^{\left( i \right)} \right),
\end{eqnarray}
where $Z$ denotes partition function, $\beta$ inverse temperature, $U$ potential energy, and summation taken over possible configurations. In alloy thermodynamics, generalized Ising model (GIM)\cite{ce} is typically employed for describing $q_{i}$ as multisite correlation functions. Then the potential energy for configuration $k$ is expressed as
\begin{eqnarray}
U^{\left( k \right)} = \sum_{j=1}^{f} \Braket{U|q_{j}} q_{j}^{\left( k \right)},
\end{eqnarray}
where $\Braket{\quad|\quad}$ denotes inner product, e.g., $\Braket{a|b}=\rho^{-1}\sum_{k}a^{\left( k \right)}\cdot b^{\left( k \right)}$ with normalization constant $\rho$. 

Under these preparations, we can introduce two $f$-dimensional vectors of $\vec{Q}_{Z}=\left(\Braket{ q_{1}}_{Z},\cdots, \Braket{ q_{f}}_{Z}\right)$ and $\vec{U}=\left(\Braket{U|q_{1}},\cdots,\Braket{U|q_{f}}\right)$, where the former represents equilibrium configuration and the latter corresponds to potential energy surface in inner product form. Eventually, the canonical average of Eq.~\eqref{eq:can} can be interpreted as a map $\phi_{\textrm{th}}$ of
\begin{eqnarray}
\phi: \vec{U} \mapsto \vec{Q}_{Z},
\end{eqnarray}
which is generally nonlinear depending on the underlying lattice: We call this nonlinear correspondence as ``canonical nonlinearity (CN)''.
Specifically, overall behavior of the CN is fully determined in terms of the configurational geometry, informed by the configurational density of states (CDOS) independent of temperature as well as energy. 
Since the CN typically exhibit complex behavior especially for alloy configurational thermodynamics, various theoretical approaches including the Metropolis algorith, multihistogram method, multicanonical ensemble and entropic sampling, have been amply proposed for accurate prediction of alloy equilibrium properties.\cite{mc1,mc2,mc3,mc4} Additionally, optimization techniques for determining interatomic many-body interactions have been developed, based on such as cross validation, genetic algorism, and regression in machine learning.\cite{cm1,cm2,cm3,cm4,cm5,cm6}
However, they do not typically address the nature of CN, which holds particularly true in terms of the configurational geometry. 

In order to overcome these problems, we have introduced two measures of the CN as a vector field on configuration space\cite{asdf,em2} and Kullback-Leibler (KL) divergence on statistical manifold,\cite{bd} respectively corresponding to local and non-local nonlinear information at a given configuration: The two measures are proposed from the perspective of the configurational geometry, i.e., they are both independent of temperature and energy, depending purely on the landscape of the CDOS. Based on these measures, thermodynamic treatment of the CN has been recently proposed,\cite{thermocn} which enables to describe the CN for \textit{multiple} configurations through thermodynamic functions that are obtained through stochastic thermodynamic transformation from the CN as stochastic evolution of the system on configuration space. 

Despite the great advantage of the proposed thermodynamics of CN (TCN), its basic property in thermodynamic equilibrium, i.e., the Gibbs state (GBS), still remains totally unclear: e.g., how the GBS relates to the CN has not been addressed so far. We here tackle this problem, and clarify (i) concrete description of the GBS, (ii) trends in GBS properties including correlation between GBSs and averaged CN for model systems, and (iii) upper and lower bounds for the averaged CN based on the GBS information. The details are shown below.

\section{Concept and Derivation}
\subsubsection*{Nonlinearity Measure and Thermodynamic Treatment}
First, we briefly explain the basic concept of nonlocal nonlinearity on statistical manifold as KL divergence of $D_{\textrm{KL}}$: This is defined at given configuration $q_{J}=\left(q_{J1},\cdots, q_{Jf}\right)$, given by\cite{ig}
\begin{eqnarray}
D_{\textrm{NOL}}^{J} = D_{\textrm{KL}}\left(P^{\textrm{E}}_{J}:P^{\textrm{G}}_{J}\right),
\end{eqnarray}
where 
\begin{eqnarray}
\label{eq:cdoss}
P^{\textrm{E}} _{J}\left(q\right)&=& z_{J}^{-1}\cdot g\left( q \right)\exp\left[ -\beta\left( {q}\cdot {V}_{J} \right) \right]   \nonumber \\
P^{\textrm{G}}_{J} \left( {q} \right)&=& \left( z_{J}^{\textrm{G}} \right)^{-1}\cdot g^{\textrm{G}}\left( {q} \right)\exp\left[ -\beta\left( {q}\cdot {V}_{J} \right) \right]  
\end{eqnarray}
respectively corresponds to canonical distribution for measuring the CN of practical and linear systems. 
Here, $g\left( {q} \right)$ corresponds to the CDOS of practical system with covariance matrix $\Gamma$, $g^{\textrm{G}}\left( {q} \right)$ corresponds to the CDOS of synthetically linear system, given by discretized multidimensional Gaussian with the same $\Gamma$, and 
\begin{eqnarray}
z_{J} &=&  \sum_{{q}}g\left({q} \right)\exp\left[ -\beta\left( {q}\cdot {V}_{J} \right) \right] \nonumber \\
V_{J} &=& \left( -\beta\cdot\Gamma \right)^{-1}\cdot {q}_{J}.
\end{eqnarray}
From the equation, we can see that $V_{J}$ acts as the artificially prepared interatomic interaction to measure the CN, where for the linear system, canonical map $\phi$ is exactly given by $-\beta\Gamma$. 
Hereafter, we employ the superscript $\textrm{G}$ as a function of the linear system, as defined for $P^{\textrm{G}}$ and $g^{\textrm{G}}$. Note that  $D^{J}_{\textrm{NOL}}$ is independent of the temperature and many-body interactions, which is a desired characteristics for configurational geometry. 

Next, we briefly explain the concept of the TCN. In the TCN, stochastic evolution of the system on configuration space, driven by the CN, is characterized by the following stochastic matrix $\mathbf{T}$:
\begin{eqnarray}
\label{eq:matT}
T_{ki} = z_{i}^{-1} g_{k} \exp\left[q_{k}\Gamma^{-1}q_{i} \right],
\end{eqnarray}
corresponding to the transition probability from configuration $q_{i}$ to $q_{k}$, where $g_{k}=g\left(q_{k}\right)$. From the definition of $\mathbf{T}$, we see that the matrix $\mathbf{T}$ naturally includes information about the CN at each configuration, since the $j$-th column of $T$ corresponds to the equilibrium distribution of $P^{\textrm{E}}_{j}$: This certainly indicates that the system transition on configuration space is driven by the CN. Based on the matrix $\mathbf{T}$, stochastic time evolution of the system is then transformed into that of thermodynamic system contacting with a thermal bath through stochastic thermodynamics, enabling to characterize CN on multiple configurations through thermodynamic functions of the linear system, which can be analytically treated in terms of the lattice geometry. For instance, 
we can provide the proper bound for the CN in terms of the entropy production for linear system $\sigma_{\textrm{G}}$, namely, 
\begin{eqnarray}
\label{eq:tcn-ex}
\Braket{\Delta D'_{\textrm{NOL}}}_{P^{+}} \le \ln \Braket{e^{-\sigma^{\textrm{G}}}}_{P^{+}}.
\end{eqnarray}
Here, $\Braket{\quad}_{P^{+}}$ denotes difference in the averaged CN between partially-ordered and other configurations: The equation denotes that averaged difference in the CN is bounded from above by $\sigma^{\textrm{G}}$, which can be exactly, analytically estimated through the covariance matrix of CDOS for practical system. As seen, the TCN has gread advantage over existing approaches, which enables to handle CN on multiple configurationssuch as in Eq.~\eqref{eq:tcn-ex}. However, its basic properties even at thermodynamic equilibrium, i.e., Gibbs state (GBS), has not been well addressed so far. In the followings, we first clarify the concrete expression of the GBS for TCN, and then discuss how the GBS relates to the information about CN. 

\subsubsection*{Gibbs State for the TCN}
We first define the GBS $\pi\in\mathbb{P}_{f}$ for TCN, which is a steady state satsfying the detailed balance condition:
\begin{eqnarray}
\label{eq:db}
T_{ij}\pi_{j} = T_{ji}\pi_{i}.
\end{eqnarray}
Since from Eq.~\eqref{eq:matT}, it is clear that $\mathbf{T}$ holds strong connectivity, given probability distribution $p$ satisfying Eq.~\eqref{eq:db} corresponds to a unique GBS. We thus express the ratio of transition probabilities as
\begin{eqnarray}
\label{eq:ratio}
\frac{T_{ij}}{T_{ji}} = \dfrac{g_{i}\exp\left(q_{i}\Gamma^{-1}q_{j}\right)z_{j}^{-1}}{g_{j}\exp\left(q_{j}\Gamma^{-1}q_{i}\right)z_{i}^{-1}} = \dfrac{g_{i}z_{i}}{g_{j}z_{j}},
\end{eqnarray}
where the last equation is obtained since $\Gamma$ is symmetric by its definition. 
From Eqs.~\eqref{eq:db} and \eqref{eq:ratio}, we can immediately obtain the concrete expression for the GBS:
\begin{eqnarray}
\label{eq:gbs}
\pi_{i} = \frac{g_{i}z_{i}}{\sum_{i}g_{i}z_{i}} = A g_{i}z_{i},
\end{eqnarray}
where $A=\left(\sum_{i}g_{i}Z_{i}\right)^{-1}$.

Based on Eq.~\eqref{eq:gbs}, we can address the basic properties of the derived GBS. Hereinafter, we take the origin of the configuration space as perfectly random state, corresponding to the 1st-order moment of the CDOS. 
We first take the ratio of CDOS and GBS at random configuration $q_{0}$: 
\begin{eqnarray}
\frac{g_{0}}{\pi_{0}} &=& \frac{1}{Az_{0}} = \Braket{\exp\left(q_{i}\Gamma^{-1}q_{j}\right)}_{g_{i}g_{j}} \nonumber \\
&\ge& \exp\left(\Braket{q_{i}}_{g_{i}}\Gamma^{-1}\Braket{q_{j}}_{g_{j}}\right) = 1,
\end{eqnarray}
where we define the average $\Braket{M_{i}}_{g_{i}}=\sum_{i}M_{i}g_{i}$.
Therefore, we get 
\begin{eqnarray}
\label{eq:gbs-p1}
g_{0} \ge \pi_{0}.
\end{eqnarray}
Next, we take the ratio of GBS at random $\pi_{0}$ and other configuration $\pi_{i}\left(i\neq 0\right)$: 
\begin{eqnarray}
\label{eq:gbs-p2}
\frac{\pi_{i}}{\pi_{0}} &=& \frac{g_{i}}{g_{0}} \Braket{\exp\left(q_{k}\Gamma^{-1}q_{i}\right)}_{g_{k}} \nonumber \\
&\ge& \frac{g_{i}}{g_{0}} \exp\left(\Braket{q_{k}}_{g_{k}}\Gamma^{-1}q_{i}\right) = \frac{g_{i}}{g_{0}}.
\end{eqnarray}
From Eqs.~\eqref{eq:gbs-p1} and \eqref{eq:gbs-p2}, and based on the fact that CDOS for alloys typically takes maxima around $q_{0}$, we can see that GBS tends to be more \textit{averaged-out} landscape from the CDOS, especially around the random configuration.

\subsubsection*{Gibbs State and Canonical Nonlinearity}
From above discussions, the role of GBS on CN is still totally unclear. Since CN is always measured from linear system, information about GBS for linear system, $\pi^{\textrm{G}}$, would be naturally required to link the GBS and CN. Therefore, we first investigate the correlation between $D_{\textrm{KL}}\left(\pi:\pi^{\textrm{G}}\right)$ and information about the CN. 

Since (i) difference in GBS between practical and linear system, $D_{\textrm{KL}}\left(\pi:\pi^{\textrm{G}}\right)$, would contain \textit{averaged} information about the CN over multiple configurations, and (ii) stochastic map $\mathbf{T}$ provides monotonic character of $D_{\textrm{KL}}\left(p:q\right)\ge D_{\textrm{KL}}\left(p':q'\right)$ where $p'=\mathbf{T}p$ and $q'=\mathbf{T}q$, we expect that the CN averaged over the GBS $\pi$ can relate to the difference in GBSs. 

To confirm the correlation between GBS and CN, we here employ 19-state CDOS with $m=1$ SDF as a model, where density of states for each state is randomly assigned: Based on the model,10,000 CDOSs are constructed. For each CDOS, we estimate (i) GBS for practical and linear system $\pi$ and $\pi^{\textrm{G}}$ through eigenvector of the individual transition matrix $\mathbf{T}$ and $\mathbf{T}^{\textrm{G}}$, and (ii) averaged CN over the GBS, $\Braket{D_{\textrm{NOL}}}_{\pi}=\Braket{D^{J}_{\textrm{NOL}}}_{\pi_{J}}$. 

\begin{figure}[h]
\begin{center}
\includegraphics[width=0.8\linewidth]{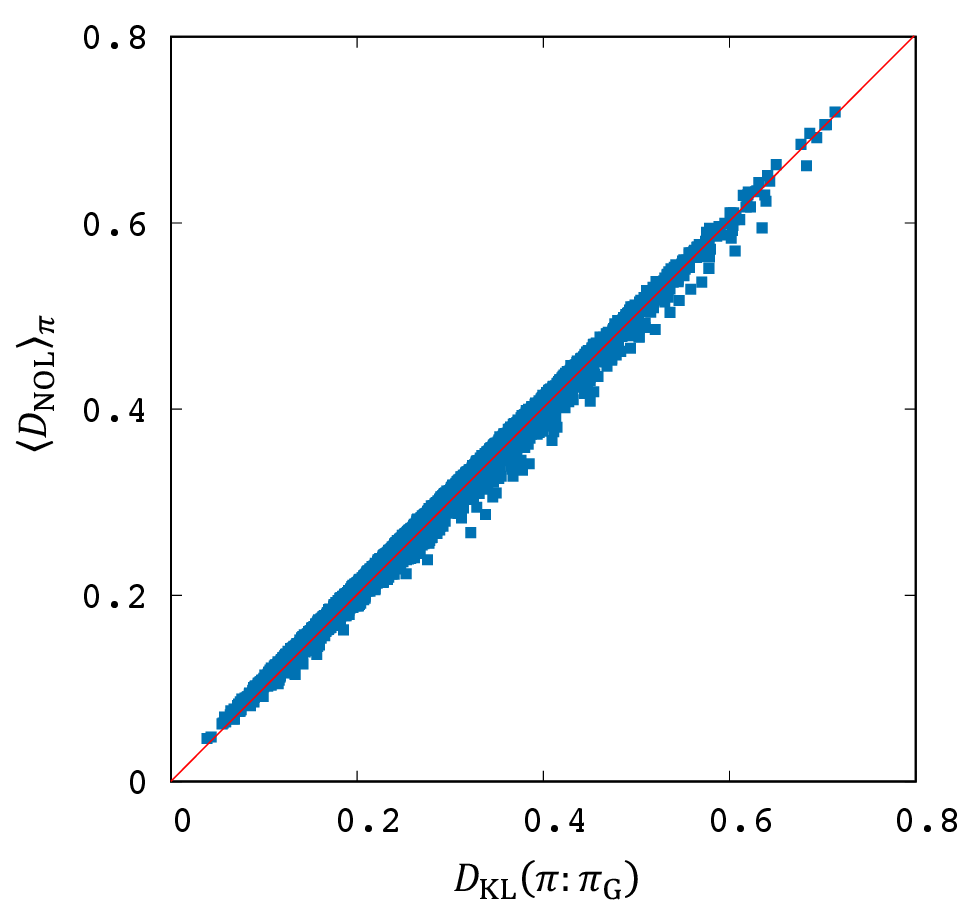}
\caption{ Correlation between $D_{\textrm{KL}}\left( \pi:\pi_{\textrm{G}} \right)$ and $\Braket{D_{\textrm{NOL}}}_{\pi}$ for 19-state CDOS models.  }
\label{fig:dkl-nol}
\end{center}
\end{figure}

We show in Fig.~\ref{fig:dkl-nol} the resultant relationshipes between $D_{\textrm{KL}}\left( \pi:\pi_{\textrm{G}} \right)$ and $\Braket{D_{\textrm{NOL}}}_{\pi}$, which cleary exhibits strong positive correlation. 
The figure strongly indicates that GBS for the TCN has profound connection to the averaged information about the nonlinearity. 

To address whether or not such connection is model-dependent, we first employ the following relationship between $\mathbf{T}$ and canonical distribution for configuration $q_{j}$:
\begin{eqnarray}
P_{J}^{\textrm{E}} = \left( T_{1j},\cdots, T_{fj} \right),
\end{eqnarray}
which therefore rewrite the CN averaged over the GBS as
\begin{eqnarray}
\label{eq:dnol-ave}
\Braket{D_{\textrm{NOL}}}_{\pi} = \Braket{ \sum_{i}T_{ij}\ln\dfrac{T_{ij}}{T^{\textrm{G}}_{ij} } }_{\pi_{j}},
\end{eqnarray}
where we define the following average:
\begin{eqnarray}
\Braket{M}_{\pi_{j}} = \sum_{j} M_{j} \pi_{j}.
\end{eqnarray}
Meanwhile,  when we consider that the GBS satisfy $\mathbf{T}\pi=\pi$, GBS is provided as
\begin{eqnarray}
\pi_{i} = \Braket{T_{ij}}_{\pi_{j}}.
\end{eqnarray}
Therefore, KL divergence of GBS between practical and linear system can be rewritten as
\begin{eqnarray}
\label{eq:nol-pi}
D_{\textrm{KL}}\left( \pi : \pi^{\textrm{G}} \right) = \sum_{i} \Braket{T_{ij}}_{\pi_{j}}\ln \dfrac{\Braket{T_{ij}}_{\pi_{j}}}{\Braket{T^{\textrm{G}}_{ij}}_{\pi^{\textrm{G}}_{j}} }. 
\end{eqnarray}
When we compare Eqs.~\eqref{eq:dnol-ave} and \eqref{eq:nol-pi}, we can qualitatively see the similarity in the formulation of $D_{\textrm{KL}}\left( \pi:\pi_{\textrm{G}} \right)$ and $\Braket{D_{\textrm{NOL}}}_{\pi}$, and also see their  difference between $D_{\textrm{KL}}\left( \pi:\pi_{\textrm{G}} \right)$ and $\Braket{D_{\textrm{NOL}}}_{\pi}$: The latter mainly comes from (i) difference in taking average $\Braket{\quad}_{\pi_{j}}$ for KL divergence itself and for $j$-th column of the transition marix $T_{ij}$ in the KL divergence, and taking average $\Braket{\quad}_{\pi^{\textrm{G}}_{j}}$ for $j$-th column of the transition matrix for linear system, $T^{\textrm{G}}_{ij}$.

\subsubsection*{ Gibbs State as Bounds for Canonical Nonlinearity}
From above discussions, we see qualitative similarity in its formulation, which does not provide sufficient information about the connection between the GBS and the CN. 
To further address this point, we here consider two distinct cases of the GBS, i.e., (i) the GBS only has a single dominant maxima at specific configuration, and (ii) the GBS takes uniform distribution of $u$. 

We first consider the former case where the GBS takes a single dominant maxima at configuration $q_{J}$, i.e., 
\begin{eqnarray}
\pi_{J} &\simeq &1 \nonumber \\
\pi_{k} &\ll& \pi_{J}\left( k\neq J \right).
\end{eqnarray}
In this condition, since $x\ln x =0\left( x\to 0 \right)$ and $\textrm{supp}\left( \pi \right) \subset \textrm{supp}\left( \pi^{\textrm{G}} \right)$, we obtain
\begin{eqnarray}
D_{\textrm{KL}}\left( \pi:\pi^{\textrm{G}} \right) \simeq \sum_{i} T_{iJ} \ln\dfrac{T_{iJ}}{T_{iJ}^{\textrm{G}}\pi^{\textrm{G}}_{J} } 
\end{eqnarray}
and
\begin{eqnarray}
\Braket{D_{\textrm{NOL}}}_{\pi} \simeq \sum_{i}T_{iJ} \ln\dfrac{T_{iJ}}{T^{\textrm{G}}_{iJ} }.
\end{eqnarray}
Therefore, we can immediately see that 
\begin{eqnarray}
D_{\textrm{KL}}\left( \pi:\pi^{\textrm{G}} \right) - \Braket{D_{\textrm{NOL}}}_{\pi} = \sum_{i} T_{iJ} \ln\dfrac{1}{\pi^{\textrm{G}}_{J} } \ge 0,
\end{eqnarray}
which means that information about GBS of practical and linear system acts as upper bound for averaged CN:
\begin{eqnarray}
\Braket{D_{\textrm{NOL}} }_{\pi} \le D_{\textrm{KL}}\left( \pi:\pi^{\textrm{G}} \right).
\end{eqnarray}

Next, we consider the latter case of $\pi=u=\left( n^{-1},\cdots, n^{-1} \right)$. Under this condition, since $\mathbf{T}$ should be doubly stochastic map satisfying
\begin{eqnarray}
\sum_{k} T_{ik} = 1,
\end{eqnarray}
we define the following probability distribution:
\begin{eqnarray}
\mathcal{P}_{i} = \left( T_{i1},\cdots, T_{in} \right) \in \mathbb{P}_{n}
\end{eqnarray}
and its corresponding shannon entropy of $S\left( \mathcal{P}_{i} \right)$. 
Based on these preparations, KL divergence for GBSs and averaged CN can be respectively rewritten as 
\begin{eqnarray}
D_{\textrm{KL}}\left( \pi:\pi^{\textrm{G}} \right) &=& -\frac{1}{n} \sum_{i} \left[ \ln n + \ln \pi_{i}^{\textrm{G}} \right] \nonumber \\
\Braket{D_{\textrm{NOL}}}_{\pi} &=& -\frac{1}{n} \sum_{i}\sum_{k} T_{ik}\ln T_{ik}^{\textrm{G}} + S\left( \mathcal{P}_{i} \right),
\end{eqnarray}
and thus, their difference is given by
\begin{widetext}
\begin{eqnarray}
\label{eq:diffd}
D_{\textrm{KL}}\left( \pi:\pi^{\textrm{G}} \right) - \Braket{D_{\textrm{NOL}}}_{\pi} = \frac{1}{n}\sum_{i}\left[ \sum_{k}T_{ik}\ln T_{ik}^{\textrm{G}} + S\left( \mathcal{P}_{i} \right) - \ln n - \ln\sum_{k}T_{ik}^{\textrm{G}}\pi_{k}^{\textrm{G}}  \right].
\end{eqnarray}
\end{widetext}
When we consider the typical character of shannon entropy of 
\begin{eqnarray}
0\le S\left( \mathcal{P}_{i} \right) \le \ln n,
\end{eqnarray}
its upper bound is achieved at $\mathcal{P}_{i}=u$. When we substitute this condition into Eq.~\eqref{eq:diffd} and employing Jensen's inequality, we get
\begin{eqnarray}
\label{eq:diffd2}
D_{\textrm{KL}}\left( \pi:\pi^{\textrm{G}} \right) - \Braket{D_{\textrm{NOL}}}_{\pi} &\le& \frac{1}{n}\sum_{i}\left[ \sum_{k} u_{k}\ln T_{ik}^{\textrm{G}} - \ln\sum_{k}T_{ik}^{\textrm{G}}\pi_{k}^{\textrm{G}} \right] \nonumber \\
&\le& \frac{1}{n}\sum_{i}\left[ \ln\sum_{k}u_{k}T_{ik}^{\textrm{G}} - \ln\sum_{k}\pi_{k}^{\textrm{G}}T_{ik}^{\textrm{G}} \right] \nonumber \\
&=& \frac{1}{n}\sum_{i}\left[ \ln\dfrac{\sum_{k}u_{k}T_{ik}^{\textrm{G}} }{\sum_{k}\pi_{k}^{\textrm{G}}T_{ik}^{\textrm{G}} } \right].
\end{eqnarray}
To further address the above difference, we here consider that $\Gamma$ is diagonal, i.e, $\Gamma=\textrm{diag}\left( \Gamma_{1}, \cdots, \Gamma_{f} \right)$, which holds for any binary alloys with pair correlations at thermodynamic limit. From Eq.~\eqref{eq:gbs}, we see that 
\begin{eqnarray}
\label{eq:gbsg}
\pi_{i}^{\textrm{G}} \propto g^{\textrm{G}}_{i}z^{\textrm{G}}_{i} = \sum_{m} g^{\textrm{G}}_{m}\exp\left( q_{m}\Gamma^{-1}q_{i} \right).
\end{eqnarray}
Under this condition, we naturally take the following continuous limit of Eq.~\eqref{eq:gbsg}:
\begin{eqnarray}
\label{eq:pigu}
g^{\textrm{G}}_{i}Z^{\textrm{G}}_{i} &=& \prod_{l=1}^{f} \int \frac{1}{2\pi \Gamma_{l}} \exp\left( -\dfrac{q_{il}^{2}+q_{ml}^{2} }{2\Gamma_{l} } \right) \exp\left(\dfrac{q_{il}q_{ml}}{\Gamma_{l} } \right) dq_{ml} \nonumber \\
&=& \prod_{l=1}^{f} \dfrac{1}{\sqrt{2\pi \Gamma_{l}}},
\end{eqnarray}
which indicates that $\pi^{\textrm{G}}=u$. Substituting Eq.~\eqref{eq:pigu} into Eq.~\eqref{eq:diffd2}, we finally obtain the lower bound for the averaged CN:
\begin{eqnarray}
\label{eq:lb}
D_{\textrm{KL}}\left( \pi:\pi^{\textrm{G}} \right) \le \Braket{D_{\textrm{NOL}}}_{\pi}.
\end{eqnarray}
Therefore, when GBS for practical system takes uniform distribution, KL divergence of GBS between practical and linear system acts as lower bound for the averaged CN. Note that from Eq.~\eqref{eq:pigu}, we see that $D\left( \pi:\pi^{\textrm{G}} \right)=0$ always holds under $\pi=u$, which means that the GBS cannot provide effective information about the averaged CN over the nature of nonegativity for the KL divergence. 

We see that the GBS at specific conditions acts as bounds for the averaged canonical nonlinearity. Considering that the initial state of CDOS for practical and linear system provides nonlocal CN at random configuration, time evolution of the system within the present thermodynamics would contain further CN information about multiple configurations other than random or GBS-weighted configurations, which will be discussed in our future study.

\section{Conclusions}
Within the thermodynamic treatment for nonlinearity in canonical ensemble, we investigate the basic properties of the GBS in terms of the configurational geometry, and reveal its role on the CN. We find the strong positive correlation between GBS for practical and linear system, and averaged CN over the GBS, which can be qualitatively understood by the similarity in their formulations. We show that at specific conditions for GBS taking a single dominant peak and uniform distribution, the GBS respectively acts as upper and lower bound for the averaged CN.

\section{Acknowledgement}
This work was supported by Grant-in-Aids for Scientific Research on Innovative Areas on High Entropy Alloys through the grant number JP18H05453 and  from the MEXT of Japan, and Research Grant from Hitachi Metals$\cdot$Materials Science Foundation.


\begin{thebibliography}{9}
\bibitem{ce} J.M. Sanchez, F. Ducastelle, and D. Gratias, Physica A \textbf{128}, 334 (1984).

\bibitem{mc1} N. Metropolis, A. W. Rosenbluth, M. N. Rosenbluth, A. H. Tellerand, and E. Teller, J. Chem. Phys. \textbf{21}, 1087 (1953).
\bibitem{mc2} A.M. Ferrenberg and R. H. Swendsen, Phys. Rev. Lett. \textbf{63}, 1195 (1989).
\bibitem{mc3} G. Bhanot, R. Salvador, S. Black, P. Carter, and R. Toral, Phys. Rev. Lett. \textbf{59}, 803 (1987).
\bibitem{mc4} J. Lee, Phys. Rev. Lett. 71, 211 (1993).

\bibitem{cm1} V. Blum, G. L. W. Hart, M. J. Walorski, and A. Zunger, Phys. Rev. B \textbf{72}, 165113 (2005).
\bibitem{cm2} A. Seko, Y. Koyama, and I. Tanaka, Phys. Rev. B \textbf{80}, 165122 (2009).
\bibitem{cm3} T. Mueller and G. Ceder, Phys. Rev. B \textbf{82}, 184107 (2010).
\bibitem{cm4} K. Yuge, Phys. Rev. B \textbf{85}, 144105 (2012).
\bibitem{cm5} L.J. Nelson, G.L.W. Hart, F. Zhou,and V. Ozolins, Phys. Rev. B \textbf{87}, 035125 (2013).
\bibitem{cm6} A.R. Natarajan and A. Van der Ven, NPJ Comput. Mater. \textbf{4}, 56 (2018).

\bibitem{asdf} K. Yuge, J. Phys. Soc. Jpn. \textbf{86}, 104802 (2018).
\bibitem{em2} K. Yuge, J. Phys. Soc. Jpn. \textbf{85}, 024802  (2016).
\bibitem{bd} K. Yuge and S. Ohta, J. Phys. Soc. Jpn. \textbf{88}, 104803 (2019).

\bibitem{thermocn} K. Yuge, J. Phys. Soc. Jpn. \textbf{93}, 094802 (2024).

\bibitem{ig} K. Yuge, J. Phys. Soc. Jpn. \textbf{91}, 014802 (2022).

\end{thebibliography}
\end{document}